\documentstyle[12pt]{article}
\newcommand{\alt}{\mathrel{\raisebox{-.6ex}{$\stackrel{\textstyle<}{\sim}$}}}
\newcommand{\agt}{\mathrel{\raisebox{-.6ex}{$\stackrel{\textstyle>}{\sim}$}}}
\def\overlay#1#2{\ifmmode \setbox 0=\hbox {$#1$}\setbox 1=\hbox to\wd 0{\hss
$#2$\hss }\else \setbox 0=\hbox {#1}\setbox 1=\hbox to\wd 0{\hss #2\hss }\fi
#1\hskip -\wd 0\box 1}
\def\case#1/#2{{\textstyle{#1\over#2}}}

\def\etal{{\it et al.}}

\catcode`@=11
\newcount\@tempcntc
\def\@citex[#1]#2{\if@filesw\immediate\write\@auxout{\string\citation{#2}}\fi
  \@tempcnta\z@\@tempcntb\m@ne\def\@citea{}\@cite{\@for\@citeb:=#2\do
    {\@ifundefined
      {b@\@citeb}{\@citeo\@tempcntb\m@ne\@citea\def\@citea{,}{\bf ?}\@warning
      {Citation `\@citeb' on page \thepage \space undefined}}%
    {\setbox\z@\hbox{\global\@tempcntc0\csname b@\@citeb\endcsname\relax}%
     \ifnum\@tempcntc=\z@ \@citeo\@tempcntb\m@ne
       \@citea\def\@citea{,}\hbox{\csname b@\@citeb\endcsname}%
     \else
      \advance\@tempcntb\@ne
      \ifnum\@tempcntb=\@tempcntc
      \else\advance\@tempcntb\m@ne\@citeo
      \@tempcnta\@tempcntc\@tempcntb\@tempcntc\fi\fi}}\@citeo}{#1}}
\def\@citeo{\ifnum\@tempcnta>\@tempcntb\else\@citea\def\@citea{,}%
 \ifnum\@tempcnta=\@tempcntb\the\@tempcnta\else
  {\advance\@tempcnta\@ne\ifnum\@tempcnta=\@tempcntb \else \def\@citea{--}\fi
   \advance\@tempcnta\m@ne\the\@tempcnta\@citea\the\@tempcntb}\fi\fi}
\catcode`@=12

\font\fortssbx=cmssbx10 scaled \magstep2

\begin{document}

\thispagestyle{empty}

\hbox to \hsize{
\hbox{\fortssbx Rutherford Appleton Laboratory}
\hfill  $\vcenter{\hbox{\bf MADPH-95-903}\vskip-.3cm
                  \hbox{\bf UICHEP-TH/95-8}\vskip-.3cm
                  \hbox{\bf RAL-TR-95-031}\vskip-.3cm
                  \hbox{\bf hep-ph/9507426}\vskip-.3cm
                  \hbox{July 1995}}$ }

\vspace{.75in}

\begin{center}
{\large\bf Possible sneutrino-pair signatures with R-parity breaking}\\[.4in]
V.~Barger$^a$, W.-Y.~Keung$^b$ and R.J.N.~Phillips$^c$\\[.2in]
\it
$^a$Physics Department, University of Wisconsin, Madison, WI 53706, USA\\
$^b$Physics Department, University of Illinois at Chicago, IL 60607-7059, USA\\
$^c$Rutherford Appleton Laboratory, Chilton, Didcot, Oxon OX11 0QX, UK
\end{center}

\vspace{.5in}

\begin{abstract}
If sneutrinos are the lightest supersymmetry partners and $R$-parity is not
conserved, the process $e^+e^- \to \tilde {\bar\nu} \tilde {\nu}$  can have
striking signatures due to the decay modes $\tilde{\nu}\to\ell^+\ell'^-$
or $\tilde{\nu}\to q \bar q'$.  We present cross section formulas and
discuss event rates and detection at the upgraded $e^+e^-$ collider LEP.
Four-lepton signals should be detectable up to sneutrino mass
$\tilde {m}_\nu = 80$ GeV and maybe beyond; four-jet signals should be
detectable up to  $\tilde {m}_\nu = 70$ GeV, but would probably be
obscured thereafter by $WW$ background.
\end{abstract}

\newpage
Searches for supersymmetry (SUSY) particles depend considerably on
the identity of the lightest SUSY partner (LSP), usually believed
to be a neutralino\cite{mssm}.
Stable sneutrinos $\tilde \nu$ are strongly disfavored as LSP
candidates, by a combination of constraints from $Z$ decay
\cite{bdt} that nowadays give mass $\tilde {m}_\nu \agt M_Z/2$ \cite{lep}
and galactic dark-matter searches that together exclude the range
$4$ GeV $< \tilde {m}_\nu < 1$ TeV \cite{caldwell,sato}. But in the
presence of $R$-parity violation (RPV)\cite{rpv,bgh,emr,dr,dl} the LSP
can be unstable and sneutrinos are credible candidates once more;
indeed the literature contains examples of SUSY-GUT parameter sets
that lead to sneutrinos as LSP, either without\cite{bbo} or
with\cite{McCurry} RPV effects in the evolution equations.
In the present paper we discuss and calculate the SUSY signals
that will arise at an $e^+e^-$ collider from the SUSY threshold
process
\begin{equation}
          e^+ e^- \to \tilde {\bar\nu}_\alpha \tilde {\nu}_\alpha \quad
                       (\alpha=e,\mu , \tau ),
\label{eq:prod}
\end{equation}
if approximately degenerate sneutrinos $\tilde{\nu}_\alpha$ are the
LSP and decay by RPV processes.  We ignore single-SUSY-particle
production\cite{bgh,dl}, that can in principle take place with a
lower threshold via RPV interactions, because the latter are either
known\cite{bgh,emr} or suspected to have much much smaller couplings
than the gauge couplings which control Eq.(\ref{eq:prod}).

Sneutrino pair production proceeds via $t$-channel exchange of
charginos $\chi_i^\pm$ (for $e_L^- e_R^+ \to \tilde {\nu}_e
\tilde {\bar {\nu}}_e$ only) and via $s$-channel Z exchange,
that together give the helicity amplitudes
\begin{eqnarray}
     {\cal M}(e_L^- e_R^+ \rightarrow \tilde {\nu}_e
                             \tilde {\bar {\nu}}_e )
&=&{e^2 \beta s \sin\theta \over 2\sin^2\theta_W}
    \left( {\cos^2\gamma_R\over t-m^2_{\chi_1^+}} +
           {\sin^2\gamma_R\over t-m^2_{\chi_2^+}} +
           {-{1\over 2}+\sin^2\theta_W \over \cos^2\theta_W(s-M_Z^2)}
    \right),
\label{eq:mlre}\\
     {\cal M}(e_L^- e_R^+ \rightarrow \tilde {\nu}_\alpha
                             \tilde {\bar {\nu}}_\alpha )
&=&{e^2 \beta s \sin\theta \over 2\sin^2\theta_W}
    \left( {-{1\over 2}+\sin^2\theta_W \over \cos^2\theta_W(s-M_Z^2)}
    \right) \quad (\alpha = \mu , \tau ),
\label{eq:mlrm}\\
{\cal M}(e_R^- e_L^+  \rightarrow\tilde{\nu}_\alpha \tilde{\bar{\nu}}_\alpha)
&=&{e^2 \beta s \sin\theta \over 2\sin^2\theta_W}
    \left( {\tan^2\theta_W \over s-M_Z^2 }
    \right) \quad (\alpha = e,\mu ,\tau ).
\label{eq:mrl}
\end{eqnarray}
Here $\theta$ is the CM polar scattering angle, $m_{\chi_i^+}$ are
the chargino masses ($i=1$ is the lightest), while
$\cos^2\gamma_R$ is the mixing probability of the wino component in
the lightest chargino;  $\beta = \sqrt {1-4\tilde{m}_\nu^2 /s}\;$
is the sneutrino CM velocity,
$s$ and $t$ are the usual invariant squares of energy and momentum
transfer, $\theta_W$ is the weak angle. The differential
cross section is then defined by $d\sigma /d\cos\theta =
\Sigma|{\cal M}|^2 \, \beta /(128\pi s)$; we note that it contains
an overall factor $\sin^2\theta$ , favoring wide angles with good
experimental acceptance, well away from the beam-pipe.
Figure 1 shows the integrated cross sections for energies
$s=140,160,175,190$ GeV, soon to be explored in successive upgrades of
the LEP collider at CERN, for a range of $\tilde{m}_\nu$ values with
$\cos^2\gamma_R \simeq 1$ and $m_\chi=2\tilde{m}_\nu$ (envisaging a
scenario with the LSP mass $\tilde {m}_\nu$ close to the lightest
neutralino mass $m_{\chi_1^0}$ and with the lightest chargino mass
$m_{\chi_1^+} \simeq 2m_{\chi_1^0}$ as in many models).
We note that $\tilde {\bar\nu}_e \tilde {\nu}_e$ production is
considerably enhanced above $\tilde {\bar\nu}_\mu \tilde {\nu}_\mu$
and $\tilde {\bar\nu}_\tau \tilde {\nu}_\tau $ production, due to the
chargino-exchange contribution (with constructive interference).
Adding all three flavors, Fig.1 indicates about 500 (75) events per
100 $pb^{-1}$ luminosity at $\sqrt s = 190$ GeV,  for
$\tilde {m}_\nu =50\,(80)$ GeV,  which should be enough to establish
a clear signal in this mass range and possibly beyond, modulo
detection efficiency and background questions discussed below.

If a sneutrino is the LSP, it can only decay by RPV. With the
particle content of the minimal SUSY-SM (MSSM), the most general
gauge- and SUSY-invariant Lagrangian includes the following terms
that can mediate sneutrino decays\cite{rpv}
\begin{equation}
{\cal L}_{RPV} = \lambda_{ijk} L_i L_j E_k^c
                +\lambda_{ijk}^\prime L_i Q_j D_k^c,
\label{eq:rpv}
\end{equation}
where $L_i$ and $E_i^c$ are the (left-handed)  lepton doublet and
antilepton singlet chiral superfields (with generation index $i$),
while $Q_i$ and $D_i^c$ are the quark doublet and charge-1/3
antiquark singlet superfields.  Antisymmetry gives $\lambda_{ijk}=-
\lambda_{jik}$ . In the MSSM these terms are all conventionally
forbidden by a multiplicative symmetry called $R$-parity ($R_p$),
with $R_p=1$ for all SM particles and $R_p=-1$ for their SUSY partners,
in order to prevent rapid proton decay.  However, proton decay is
forbidden if there are no additional $B$-violating terms, in which case
either or both classes of $L$-violating terms above are allowed.
Each such RPV term provides a possible decay channel into SM fermions
as follows:
\begin{eqnarray}
\lambda_{ijk}
&\Rightarrow&\tilde{\nu}_i\to\ell_{jR}^+\ell_{kR}^-,\quad
             \tilde{\nu}_j\to\ell_{iR}^+\ell_{kR}^-,
\label{eq:dec1}\\
\lambda_{ijk}^\prime&\Rightarrow&\tilde {\nu}_i\to \bar d_{jR} d_{kR},
\label{eq:dec2}
\end{eqnarray}
together with the charge-conjugate channels. The corresponding decay
widths are therefore
\begin{equation}
\Gamma(\tilde{\nu}_i\to\ell_j^+\ell_k^-) =
\lambda_{ijk}^2 \tilde {m}_\nu /(16\pi),\quad
\Gamma(\tilde{\nu}_i \to \bar d_j d_k) =
3\lambda^{'2}_{ijk} \tilde {m}_\nu /(16\pi),
\label{eq:gam}
\end{equation}
neglecting the lepton and quark masses. The requirement that
$\tilde {\nu}$ decays within the detector (typically within 1 m)
translates into
\begin{equation}
\lambda^2\; (or\;3 \lambda'^2)\; \agt \;
       \beta\gamma\; ({\rm GeV}/ \tilde {m}_\nu )\times 10^{-14},
\label{eq:life}
\end{equation}
where $\beta\gamma = \sqrt {s/(4 \tilde {m}_\nu^2) -1}$ is the
appropriate sneutrino Lorentz factor and $\lambda$ (or $\lambda'$)
denotes the coupling of the dominant decay process in Eq.(\ref{eq:dec1})
(or Eq.(\ref{eq:dec2})).  For the energies and masses of present interest,
where Fig.1 shows appreciable cross sections,
this implies very weak lower bounds $|\lambda |,|\lambda' | \agt 10^{-8}$
on the dominant couplings.  We shall assume that all sneutrino flavors
are near-degenerate and all decay directly by one of the RPV couplings
above, neglecting $R_p$-conserving decays of heavier sneutrinos to
the lightest sneutrino that are suppressed by phase-space and loop
factors [but the latter decays would have very similar signatures
anyway, with very little extra activity from the cascade process].

The alternative decay modes of Eqs.(\ref{eq:dec1} -\ref{eq:dec2}) give
quite distinctive signatures, that we now discuss.\\
(a) $LLE^c$-mediated decays: Eq.(\ref{eq:dec1}). Each sneutrino decays
to two charged leptons, not necessarily of the same flavor; e.g. the
coupling $\lambda_{121}$ would give
\begin{eqnarray}
\tilde {\nu}_e\;\to\;\mu^+e^-,\quad &
        e^+e^-\;\to\;    \tilde {\bar \nu}_e\tilde {\nu}_e\;
                \to\; & (\mu^-e^+)(\mu^+e^-),
\label{eq:emu}\\
\tilde {\nu}_\mu\;\to\;e^+e^-,\quad &
        e^+e^-\;\to\;    \tilde {\bar \nu}_\mu \tilde {\nu}_\mu\;
                \to\; & (e^-e^+)(e^+e^-),
\label{eq:ee}
\end{eqnarray}
Note that no invisible $\tilde {\nu}_i\to\nu_j\nu_k$ modes are
accessed at tree level.
Four-lepton final states like Eq.(\ref{eq:emu}), with no missing energy
and two invariant-mass constraints $m(\mu^-e^+) = m(\mu^+e^-) =
\tilde {m}_{\nu e}$ would be very striking and easily separated from SM
backgrounds, that are small of order $\alpha^4$ and mostly contain
low-mass QED pairs; the sum of $e^+e^-\to ZZ\to (\ell\ell)(\ell\ell)$
backgrounds is $\alt 10^{-2}pb$.  Final states like Eq.(\ref{eq:ee})
would be similarly constrained, but with an ambiguity in the $e^+e^-$
pairing to be resolved by mass matching.  Analogous final states
containing two $\tau$ leptons could be identified (including the
$\tau^\pm$ charges) with good efficiency from the narrow
few-prong $\tau$-decay topologies and/or displaced decay vertices.
The directions of the tau momentum vectors would be approximately
measurable and their magnitudes could be reconstructed from overall
energy and momentum conservation\cite{taus}.
Possible decays to four taus would also be recognizable and striking,
but their reconstruction would be a zero-constraint fit with no
protection against initial-state radiation corrections.
In practice, in cases where the lepton pairing is ambiguous, one
can select the pairing for which the two masses agree most closely
and define their mean to be the best-fit sneutrino mass
$\tilde {m}_\nu (\ell^+\ell^-)$.
\\
(b) $LQD^c$-mediated decays: Eq.(\ref{eq:dec2}). Each sneutrino decays
to two charge-1/3 quarks, not necessarily of the same flavor, normally
giving two jets. Flavor-tagging is possible for $b$-jets, but otherwise
these modes are all essentially indistinguishable;
\begin{equation}
\tilde {\nu}_\alpha\;\to\;jj,\quad
        e^+e^-\;\to\;    \tilde {\bar \nu}_\alpha \tilde {\nu}_\alpha\;
                \to\;  (j_1j_2)(j_3j_4),
\label{eq:jj}
\end{equation}
The 4-jet final state of Eq.(\ref{eq:jj}) has no missing energy
(except from semileptonic decays within b-jets) and two invariant-mass
constraints $m(j_1j_2) = m(j_3j_4) =\tilde {m}_{\nu\alpha}$ to
distinguish it from background.
Of three possible dijet pairings, we should
select the one where the dijet masses agree most closely, and define
their mean to be the best-fit sneutrino mass $\tilde {m}_\nu (jj)$.
QCD four-jet backgrounds are expected to be of order
$\alpha_s^2\,\times\,\sigma(e^+e^-\to \bar qq)\sim 1$ pb, of which
only a small fraction will accidentally satisfy the dijet mass
constraint.   Backgrounds from $e^+e^-\to WW\to jjjj$
are bigger, rising from about 1.5 to 9 pb over the range
$\sqrt s=160-190$ GeV, but they obey their own dijet mass constraint
$m(j_1j_2)\simeq m(j_3j_4) \simeq M_W$ that can be used to identify
and remove most of these events; however, they will obscure any
sneutrino signal with $\tilde {m}_\nu$ approaching $M_W$.
It could be be advantageous to study decays like Eq.(\ref{eq:jj})
at or below $\sqrt s=160$ GeV where the $WW$ background is relatively
suppressed compared to lower-mass sneutrino signals.  If the
sneutrino signal happens to contain a $b$-jet, then $b$-tagging
would also suppress the $WW$ background (though a fraction of the
much smaller $e^+e^-\to ZZ$ background would survive).
\\
(c) Mixed $LLE^c/LQD^c$ decays. If the leading $LLE^c$ and $LQD^c$
decay modes have comparable rates, there will be some events where
one sneutrino decays to two leptons while the other decays to two
jets, giving
\begin{equation}
        e^+e^-\;\to\;    \tilde {\bar \nu}_\alpha \tilde {\nu}_\alpha\;
                \to\; (\ell^+\ell^{'-})(j_1j_2),
\label{eq:lljj}
\end{equation}
with no missing energy (except sometimes with b-jets) and two
invariant-mass constraints $m(\ell^+\ell^{'-}) = m(j_3j_4) =
\tilde {m}_{\nu\alpha}$ to distinguish them from backgrounds,
that are small anyway (the sum of $e^+e^-\to ZZ\to (\ell\ell )(jj)$
backgrounds is $\alt 0.1$ pb).
Here $\ell$ and $\ell'$ denote $e,\;\mu$, or $\tau$.   Cases
with different lepton flavors would be especially striking and
background-free.  Although dilepton and dijet masses should
agree within resolution, the former usually gives a sharper
estimate of $\tilde {m}_\nu$.
\\
(d) Displaced vertices.  If the sneutrino mean decay length is of
order 0.1mm--1m (typically $\lambda \sim 10^{-6}-10^{-8}$ for cases
of present interest), the two sneutrino decay vertices will usually
be detectably displaced from each other and from the beam-intersection
spot, providing an important extra signature (modulo some complications
in events with final taus).  This signature would discriminate strongly
against most SM backgrounds, including $WW\to 4j$.

To illustrate the invariant mass distributions, we impose typical
gaussian resolution smearing on energies $E$, with
$\Delta E=0.2\sqrt {E/GeV}$ for leptons and
$\Delta E=0.8\sqrt {E/GeV}$ for jets.  We also impose semi-realistic
cuts, requiring all leptons and quarks to have rapidities $|\eta|<2$,
energies $E>8$ GeV, and angular separations $\theta_{ij} > 20^\circ$.
These cuts give about $80\%$ acceptance for the examples we show.
In the four-lepton and two-lepton-two-jet channels, there is little
or no ambiguity about the pairings and the resulting
best-fit sneutrino mass distributions have clean narrow peaks,
that it is unnecessary to illustrate explicitly.
The four-jet channels however have potentially serious $WW$
backgrounds.  Figure 2 illustrates the distribution of best-fit
sneutrino mass $\tilde {m}_\nu (jj)$ in four-jet cases like
Eq.(\ref{eq:jj}), for $\sqrt s=175$ GeV and $\tilde {m}_{\nu e} =
50,65,80$ GeV.  The $WW$ background is calculated from the Pythia
Monte Carlo, with a correction for missing neutrinos
from semileptonic $c$-decays in addition to the energy smearing
and acceptance cuts above.  We see that the
$\tilde {m}_\nu =50$ GeV mass peak is well above background.
Assuming integrated luminosity 100 $pb^-1$ and summing all three
sneutrino flavors, the $\tilde {m}_\nu =65$ GeV case would
predict a signal $S\simeq 100$ events in the 60-70 GeV mass bin
compared to a WW background $B \simeq 60$ events, giving
significance $S/\sqrt B = 13$. The  $\tilde {m}_\nu =70$ GeV
case would give  only $S\simeq 60$ with higher $B\simeq 120$
and lower significance $S/\sqrt B =5$; this mass is about the limit
for establishing a signal in the four-jet channel at $\sqrt s=175$
GeV.  Significance would be slightly improved with
$\sqrt s=190$ GeV instead, but high luminosity may be harder to
achieve here.

If a $\tilde {\nu}_\mu\to e^+e^-$ or $\tilde {\nu}_\tau\to
e^+e^-$ decay signal were to be established (via $\lambda_{121}$ or
$\lambda_{131}$ coupling), it would imply a
corresponding resonance\cite{bgh} in the $e^+e^-\to
\tilde {\nu} \to e^+e^-$ channel -- or indeed for
$e^+e^-\to\tilde {\nu} \to \mu^+\mu^-$ if both $\lambda_{131}$ and
$\lambda_{232}$ were significant.  Such a resonance would have a
large peak cross section of order $4\pi /\tilde {m}_\nu^2$, four
orders of magnitude above $\sigma_{QED}(e^+e^- \to \mu^+\mu^-)$,
but a very narrow width; for the case $\tilde {m}_\nu = 50$ GeV, the
upper limits\cite{bgh} $\lambda_{121} \alt 0.04(m_{\tilde {e}}/100\,GeV)$
and $\lambda_{131} \alt 0.1(m_{\tilde {e}}/100\,GeV)$ indicate
a width of 10 MeV at best, and it could be very much smaller.  If
these couplings were indeed near their upper limits above, such a
resonance could be detected by a suitable scan; indeed, the absence of
corrections to Bhabha scattering at TRISTAN already imposes
significant further limits\cite{bgh} for $\tilde {m}_\nu = 50-56$
GeV. But for smaller couplings, $\lambda_{ijk} < 10^{-3}$
say, the width would be less than 1 keV giving five orders of
magnitude suppression in a scanning bin of width 100 MeV, and the
resonance signal would be lost.  Similarly a hadronic decay signal
$\tilde {\nu}\to \bar qq'$ would imply the presence of
$\bar qq' \to \tilde {\nu}\to \bar qq'$ resonance contributions
in hadron collisions, but their narrow widths would make them
much harder to detect than the SM $\bar qq \to Z \to \bar qq$ signals,
which are themselves quite difficult to detect.

Finally, we note that hadron colliders too can put constraints on our
sneutrino=LSP scenario, not only through direct electroweak production
of sneutrinos but more importantly by hadroproduction of squarks and
gluinos that would decay eventually to LSP pairs and hence to
components like Eq.(\ref{eq:dec1}) in each final state.  Since no
anomalous four-lepton production has been reported at hadron colliders,
it would appear either that $\tilde {\nu}\to e^+e^-, e^{\pm}\mu^{\mp},
\mu^+\mu^-$ are not dominant, or that squark and gluino production
is strongly suppressed by high mass thresholds. On the other hand,
decays   $\tilde {\nu}\to e^{\pm}\tau^{\mp},\mu^{\pm}\tau^{\mp}$
with at least one tau lepton are not disfavored and neither
are analogous hadronic decays like Eq.(\ref{eq:dec2}).

Our results may be summarized as follows.\\
(1) We have shown that sneutrino pair production with RPV decays can
give substantial and distinctive four-lepton or four-jet and possibly
dilepton-dijet signals at the future upgraded LEP collider, if
the LSP is a sneutrino and sneutrinos have approximately degenerate
mass $\tilde {m}_\nu \alt 80$ GeV. \\
(2) Production of $\tilde\nu_e$ pairs gives the biggest signals;
$\tilde\nu_\mu$ and $\tilde\nu_\tau$ signals
are similar but could have different lepton/jet flavors (if all
sneutrinos decay directly via RPV) or could be essentially the same
(if heavier sneutrinos decay first to the LSP, with negligible
emission of soft particles).  All sneutrino signals will generally
be indistinguishable in the four-jet modes, in the absence of
jet-flavor tagging.\\
(3) All signals have narrow dilepton and/or dijet invariant mass
peaks at $m(\ell^+\ell^-),\,m(jj)=\tilde {m}_\nu$, and may also
possess displaced-decay-vertex signatures.\\
(4) The signals are cleanest in the four-lepton
channels ($LLE^c$-dominated decays), where there is little SM
background.  In the four-jet channel the most serious background
appears to be $WW$ production,  which is tolerable for
$\tilde {m}_\nu \alt 70$ GeV but obliterates signals near
$\tilde {m}_\nu=M_W$ (unless additional $b$-tag or displaced-vertex
signatures are present).\\
(5) If a $\tilde {\nu}_\mu\to e^+e^-$ or $\tilde {\nu}_\tau\to
e^+e^-$ decay signal were established, the corresponding
$e^+e^-\to \tilde {\nu}$ resonance signals\cite{bgh} could be
detectable via a dedicated scan at lower energy, but only if its
coupling $\lambda$ were not too far from the present upper limit.
Possible $\bar qq' \to \tilde {\nu} \to \bar qq'$ signals
would be unfeasible to detect at hadron colliders.\\
(6) The absence of reported $\ell_1\ell_2\ell_3\ell_4$ signals
($\ell_i=e,\mu$) at hadron colliders, however, suggests that
$\tilde {\nu}\to ee,e\mu,\mu\mu$ decays are unimportant, unless
SUSY hadroproduction is suppressed by high mass thresholds.

\begin{flushleft}{\bf Acknowledgments}\end{flushleft}
We are grateful to Robert Sekulin for providing samples of
$e^+e^-\to WW \to 4j$ Monte Carlo events.  RJNP thanks Gian Gopal
for helpful conversations about LEP experimentation.
This research was supported in part by the U.S.~Department of Energy
under Grants No.~DE-FG02-95ER40896 and No.~DE-FG02-84ER40173 and
in part by the University of Wisconsin Research Committee with funds
granted by the Wisconsin Alumni Research Foundation.


\section*{Figures}
\begin{enumerate}
\item{Integrated cross sections $\sigma (e^+e^-\to \tilde{\bar\nu}_\alpha
\tilde {\nu}_\alpha)$ in pb versus sneutrino mass $\tilde {m}_\nu$
for a range of LEP energies and different sneutrino flavors.
For $\tilde {\nu}_e$ pair production we assume the lightest chargino has
mass $m_\chi = 2\tilde {m}_\nu$ and mixing $\cos^2\gamma_R\simeq 1$.
\label{fig:fig1}}
\item{Best-fit dijet sneutrino mass distributions for $\tilde {\nu}_e$
production at $\sqrt s =175$ GeV with $\tilde {m}_{\nu e}=50,\;65,\;80$ GeV
and four-jet final states like Eq.(\ref{eq:jj}), after the resolution
smearing and acceptance cuts described in the text.  The principal
background from $WW\to 4j$ is shown for comparison.
\label{fig:fig2}}
\end{enumerate}

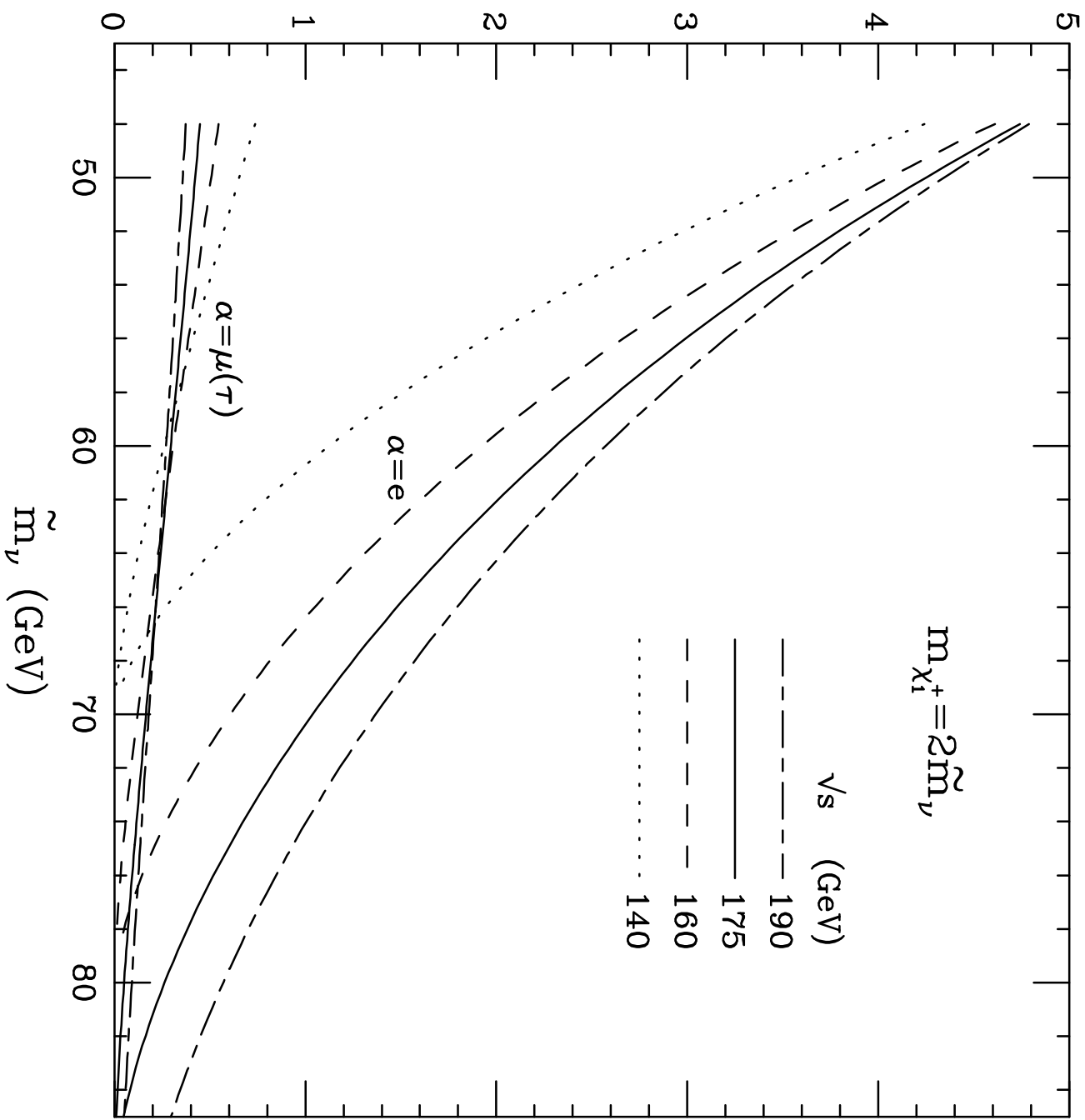
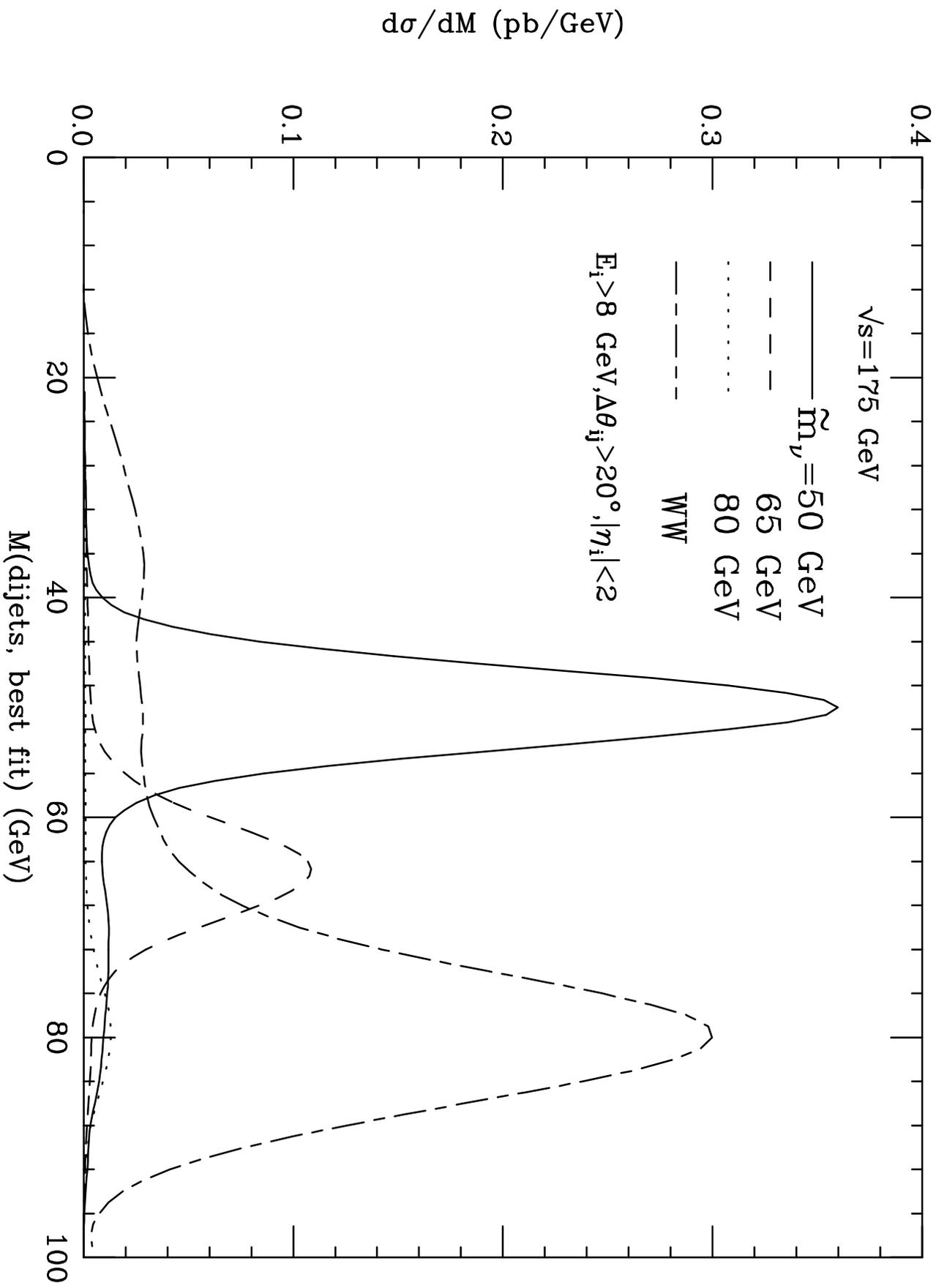
\end{document}